\documentclass[english,longbibliography,twocolumn,superscriptaddress,amsmath,amssymb]{revtex4-1}
\usepackage[T1]{fontenc}
\usepackage[latin9]{inputenc}
\usepackage{float}
\usepackage{graphicx,color}
\usepackage{subfigure}
\usepackage{epstopdf} 
\makeatletter
\@ifundefined{textcolor}{}
{%
 \definecolor{BLACK}{gray}{0}
 \definecolor{WHITE}{gray}{1}
 \definecolor{RED}{rgb}{1,0,0}
 \definecolor{GREEN}{rgb}{0,1,0}
 \definecolor{darkgreen}{rgb}{0,0.6,0}
 \definecolor{BLUE}{rgb}{0,0,1}
 \definecolor{CYAN}{cmyk}{1,0,0,0}
 \definecolor{MAGENTA}{cmyk}{0,1,0,0}
 \definecolor{YELLOW}{cmyk}{0,0,1,0}
 }
\usepackage{comment}
\newcommand{\norm}[1]{\left\lVert#1\right\rVert}

\newcommand{\figurewidth}{\columnwidth}
\newcommand{\beq}{\begin{equation}}
\newcommand{\eeq}{\end{equation}}
\newcommand{\bea}{\begin{eqnarray}}
\newcommand{\eea}{\end{eqnarray}}

\newcommand{\e}{{\text e}}

\usepackage{amsthm}

\makeatother

\usepackage{babel}

\makeatother

\usepackage{babel}
\usepackage{graphicx}
\usepackage{amsfonts}
\usepackage{amsmath}

\makeatother

\usepackage{babel}

\begin{document}

\title{Equation Planting: A Tool for Benchmarking Ising Machines}

\author{Itay Hen}
\affiliation{Information Sciences Institute, University of Southern California, Marina del Rey, California 90292, USA}
\affiliation{Department of Physics and Astronomy and Center for Quantum Information Science \& Technology, University of Southern California, Los Angeles, California 90089, USA}
\date{\today}

\begin{abstract}
\noindent  We introduce a methodology for generating benchmark problem sets for Ising machines---devices designed to solve discrete optimization problems cast as Ising models. In our approach, linear systems of equations are cast as Ising cost functions. While linear systems are easily solvable, the corresponding optimization problems are known to exhibit some of the salient features of NP-hardness, such as strong exponential scaling of heuristic solvers' runtimes and extensive distances between ground and low-lying excited states. We show how the proposed technique, which we refer to as `equation planting,' can serve as a useful tool for evaluating the utility of Ising solvers functioning either as optimizers or as ground-state samplers. We further argue that equation-planted problems can be used to probe the mechanisms underlying the operation of Ising machines. 
\end{abstract}

\maketitle

\noindent{\bf Introduction.}--- 
Recent years have witnessed a flourishing in experimental `Ising machines'---special-purpose programmable devices engineered to solve discrete optimization problems~\cite{korte2012combinatorial,papadimitriou2013combinatorial} cast as Ising models---with the tacit promise that their performance is superior to those of algorithms running on standard computers. Analog quantum devices that perform quantum annealing~\cite{kadawoki:98,farhi_long:01} designed to find bit assignments that minimize the cost of Ising Hamiltonians have already been realized on various  platforms such as arrays of superconducting flux qubits~\cite{LL1,LL2,LL3,Johnson:2010ys,Berkley:2010zr,Harris:2010kx,Bunyk:2014hb}. 
Other notable technologies that recently emerged are coherent Ising machines based on lasers and degenerate optical parametric oscillators~\cite{coherentIsing1,coherentIsing2}, FPGA-based quantum-inspired digital annealers~\cite{fujitsu}, and memcomputing devices that operate on terminal-agnostic self-organizing logic gates~\cite{memComputing1,memComputing2,memComputing3}. 

Touting improved performance over traditional algorithms~\cite{2058-9565-2-4-044002,Inagaki603}, Ising machines have gained a large amount of interest, both in the academe as well as with the general public---and rightfully so. Many problems of theoretical and practical relevance, in areas as diverse as machine learning, materials design, software verification, portfolio management, logistics, and many more, can be cast as searching for the global minima of Ising cost functions~\cite{korte2012combinatorial,papadimitriou2013combinatorial}. 
It is, however, unclear to date whether indeed any one of the aforementioned devices truly offers genuine advantages over its competitors. 

One of the main bottlenecks preventing meaningful benchmarking of Ising machines is the absence of appropriate benchmark problems. Generating problem instances that are on the one hand challenging enough and whose solutions are  on the other hand known in advance or are easily verifiable, so as to allow proper benchmarking, are not straightforward to construct since these two requirements are, in many respects, contradictory. To overcome this obstacle, various methods have been devised in recent years to generate problem classes with known minimizing configurations (see, e.g., Refs~\cite{hen:planted,King2017,PhysRevA.94.012320}). However these have generally been found to lack the hardness that characterizes NP-hard problems. At the other extreme, problem classes that are challenging to solve but whose ground-state energies are not verifiable, have also been developed~\cite{Mandr__2018, martin-mayor:15, VenturelliPRX}.

To bridge the above gap, in this study we advocate a methodology for generating benchmark problem sets for Ising machines that we argue may serve as a suitable tool for evaluating their performance and in turn also to indirectly probe the mechanisms underlying their operation. The proposed technique, which we refer to as `equation planting,' has several desirable properties. Explicitly, the generated problem sets, while generated from easily solvable problems, possess some of the salient features of NP-hard problems---most notably extensive distances between ground and low-lying excited states. 

Our approach is based on the embedding of linear systems of equations in Ising models, motivated by the observation that when linear systems of equations are cast as optimization problems, the latter form often stymies heuristic solvers~\cite{franz:01,PhysRevLett.88.188701,jorg:10,Ricci-Tersenghi1639,guidetti:11,farhi:12}. In addition to serving as suitable benchmark problems to Ising machines functioning as optimizers, we also show that the proposed technique may be used to test the functionality of these devices as Boltzmann samplers, or more specifically as ground-state samplers~\cite{adachi,perdomo, 2xdwave}, thanks to another property of these problem sets---namely, a verifiable number of ground state configurations. 

We begin by describing the equation planting technique in general, moving on to focus on a specific class of problem instances using which we demonstrate the effectiveness of the method. 

\noindent{\bf Equation planting.}--- 
We begin by considering a linear system of $m$ equations in $n$ variables
\beq
\sum_{j=1}^{n} a_{ij} x_j  = b_i  \quad \text{for} \quad i=1\ldots m \,.
\eeq
Here $\{ x_1,\ldots, x_n\}$ stand for variables over a given field, and $\{a_{ij}\}$ and $\{b_i\}$ are the equation coefficients. 
Every such linear system may be cast as an optimization problem with the corresponding cost function
\beq
F = \sum_{i=1}^{m} \norm{\sum_{j=1}^n a_{ij} x_j - b_i}^2 \,.
\eeq
Since $F$ is a sum over positive terms, any configuration $\{x^*_1,\ldots,x^*_n\}$ that yields $F=0$ is a minimizing configuration that also solves the linear system.  

As already mentioned, even though the computational problem of solving a linear system of equations is
easy---any given instance can always be solved in polynomial
time using Gaussian elimination---the corresponding optimization problem is not necessarily easy for heuristic solvers~\cite{franz:01,PhysRevLett.88.188701,Ricci-Tersenghi1639,guidetti:11,farhi:12}. 

We leverage the above setup towards constructing Ising problems that stymie heuristic solvers. To that aim, we focus henceforth on linear systems of equations modulo 2, that is, 
\beq\label{eq:xor1}
\sum_{j=1}^{n} a_{ij} x_j  = b_i  \mod 2 \quad \text{for} \quad i=1\ldots m \,,
\eeq
where  now both the variable set and coefficients are Booleans taking on values $\{0,1\}$. 
The $i$th equation can therefore be written as \hbox{$\bigoplus_{j: a_{ij}=1}{x_j} = b_i$} (here $\oplus$ denotes the bitwise XOR operation), or
in terms of Ising spins $s_i \in \{-1,1\}$,
\beq\label{eq:xor2}
\prod_{j: a_{ij}= 1} s_j = (-1)^{b_i} \,.
\eeq
Cast in optimization form, the system of equations becomes
\beq
\tilde{F}_2= \sum_{i=1}^{m} \left( \prod_{j: a_{ij}= 1} s_j - (-1)^{b_i}\right)^2  \,,
\eeq
or, after dropping immaterial constants, 
\beq \label{eq:F2}
F_2= -\sum_{i=1}^{m} (-1)^{b_i} \prod_{j: a_{ij}= 1} s_j \,.
\eeq
The cost function $F_2$ is a multi-spin cost function consisting of a sum of products of spin variables. The localities of the terms comprising the cost function 
are precisely the number of variables in the corresponding equations. 
The minima of $F_2$ correspond to the solutions of the system given in Eq.~(\ref{eq:xor2}), provided that solutions exist.  In more detail, linear systems may behave in one of three possible ways. (i) The system may have no solutions at all. This scenario corresponds to the cost function $\tilde{F}_2$ having a strictly positive minimal value (meaning $F_2>-m$) and may happen if the number of equations is greater than the number of variables. (ii) The system may have a unique solution. Here, a single configuration minimizes $F_2$ whose minimal value in this case would be $-m$. (iii) Last, the system may have multiple solutions if the dimension of its null space, or nullity~\cite{linear}, which we denote here by $d_N$, is nonzero. In this case, the number of minimizing configurations that yield $F_2=-m$ is 
$N_{\text{GS}}=2^{d_N}$. 

Ising machines are designed to tackle two-body cost functions. 
The locality of the multi-spin cost $F_2$ may in this case be readily reduced to a two-body Ising model of the general form
$\sum_{ij} J_{ij} s_i s_j + \sum_i h_i s_i$ using standard reduction techniques (see, e.g., Ref.~\cite{doi:10.1002/andp.201300120}), where $\{ J_{ij}\}$ and $\{h_i\}$ are integer-valued coefficients. 

As we demonstrate in the next section, while the solutions of the linear systems of equations used to generate these Ising cost functions are straightforwardly obtained (if there are any), heuristic solvers will generically find this type of problems extremely difficult to solve. 
In what follows, we illustrate this by studying in detail one specific type of equation system---namely, random 3-regular 3-XORSAT. 

\noindent{\bf 3-regular 3-XORSAT as 2-body Ising.}---
To demonstrate the challenges that equation-planted instances present to heuristic Ising solvers, we consider as a test case linear systems $\mod 2$ (also known as XORSAT equations) wherein each equation contains exactly three randomly chosen Boolean variables and each variable, or bit, $x_j$ appears in exactly three equations. This type of problem is also known as `3-regular 3-XORSAT' (3R3X) and was studied previously in various contexts~\cite{jorg:10, guidetti:11,farhi:12,tempScalingLaw,analogErrors}.
An $n$-bit instance would thus consist of $n$ equations of the form \hbox{ $x_{i_1}\oplus x_{i_2} \oplus x_{i_3}=b_i$}(see the Appendix for additional details). We note that while the results presented below pertain to 3R3X, they should extend in general to other classes of linear XORSAT systems.  

\begin{figure}[h]
\begin{center}
\includegraphics[width=0.5\figurewidth]{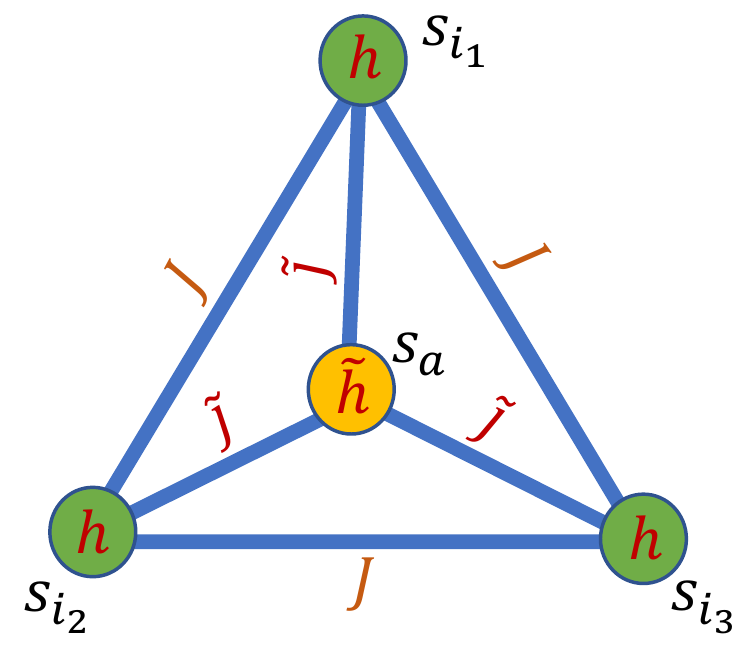}
\caption{The two-body 3-XORSAT gadget. The gadget is a fully connected spin quadruplet with parameter values $(h,\tilde{h},J,\tilde{J})$  as described in the main text. The four minimizing configurations of the 3-XORSAT clause are encoded into the ground states of the Ising cost of the gadget restricted to spins $s_{i_1},s_{i_2},s_{i_3}$. The spin $s_a$ serves as an auxiliary spin. }
\label{fig:xgadget}
\vspace{-0.7cm}
\end{center}
\end{figure}

The cost function $F_2$ of an $n$-bit 3R3X system is therefore a sum of $n$ three-body terms of the form $-(-1)^{b_i} s_{i_1} s_{i_2} s_{i_3}$ defined on $n$ Ising spins. 
To reduce the locality of a term to two- and one-body we use a gadget that shares its four minimizing configurations (see the Appendix for additional details). For the negatively signed clause ($b_i=0$), these are the four configurations whose product is positive, namely, $(1,1,1),(1,-1,-1),(-1,1,-1)$ and $(-1,-1,1)$ and an appropriate gadget is
\bea
G_{\text{3X}}&=&h(s_{i_1}+s_{i_2}+s_{i_3}) + \tilde{h} s_a \\\nonumber
&+&J(s_{i_1} s_{i_2}+s_{i_2} s_{i_3}+s_{i_3} s_{i_1})+\tilde{J} s_a(s_{i_1} +s_{i_2} +s_{i_3}) \,,
\eea
where $(h,\tilde{h},J,\tilde{J})$ can be either $(-1, -2, 1, 2)$ or $(-1, 2, 1, -2)$, yielding in both cases a minimal cost of $-4$ (we note that other gadgets exist that can equivalently be used). The above gadget is a fully connected 4-spin cost function with $s_a$ serving as an auxiliary spin (see Fig.~\ref{fig:xgadget}). The positively signed clauses are treated similarly.

The above gadget allows us to cast $n$-bit 3R3X linear systems as two-body Ising models with $2n$ spins, as each clause adds one auxiliary spin to the problem and an $n$-bit instance has $n$ clauses.

At this point, we note several of the 3R3X Ising instances properties: (i) First, these instances have a bounded degree, namely 9, as each spin connects in general to three other spins for every clause in which the spin appears. (ii) The specification of Ising parameters $\{J_{ij}\}$ and $\{h_i\}$ requires only two bits of precision as both spin-spin couplings and longitudinal fields are integers in the range $[-3,3]$. (iii) As already mentioned, these instances have known ground-state energies. (iv) Finally, these instances have a controllable ground-state degeneracy, determined by the nullity of the generating linear system. 
As will become evident in the next section, these  properties may be leveraged to determine the utility of heuristic Ising solvers functioning either as optimizers or as ground-state samplers. 

\noindent{\bf Results.}---
In what follows, we examine the performance of heuristic solvers on randomly generated two-body Ising 3R3X instances that have unique solutions. 
As a representative for state-of-the-art heuristic solvers, we use parallel tempering (PT)~\cite{hukushima:96,marinari:98b} --- a refinement of
the celebrated yet somewhat outdated simulated annealing
algorithm~\cite{kirkpatrick:83}, that finds the ground-state configurations of general discrete-variable cost functions~\cite{Tchaos1,fernandez:09,fernandez:09b,PhysRevB.62.14237,Katzgraber_2006,B509983H,martin-mayor:15,Zhang17,hen:planted,tempScalingLaw}.
In PT, multiple copies of the problem are equilibrated in parallel at different temperatures and spin configurations at adjacent temperatures are regularly swapped (see the Appendix for additional details).
In addition, we consider a variant of PT (which we denote PT-H) that employs global cluster moves due to Houdayer~\cite{Houdayer} and is known to accelerate PT convergence rates for many problem classes (see, e.g., Refs.~\cite{PhysRevB.70.014417,PhysRevB.70.014418,PhysRevX.8.031016}).

We test the performance of PT and PT-H on instances of varying sizes by measuring their typical times to find minimizing configurations.
Here, typical runtimes are defined as the median time to reach a minimizing configuration over 100 randomly generated instances of a given size~\footnote{Both algorithms were run on one core of a $3.5$ GHz 6-Core Intel Xeon E5 processor.}. 

\begin{figure}[h]
\begin{center}
\includegraphics[width=0.9\figurewidth]{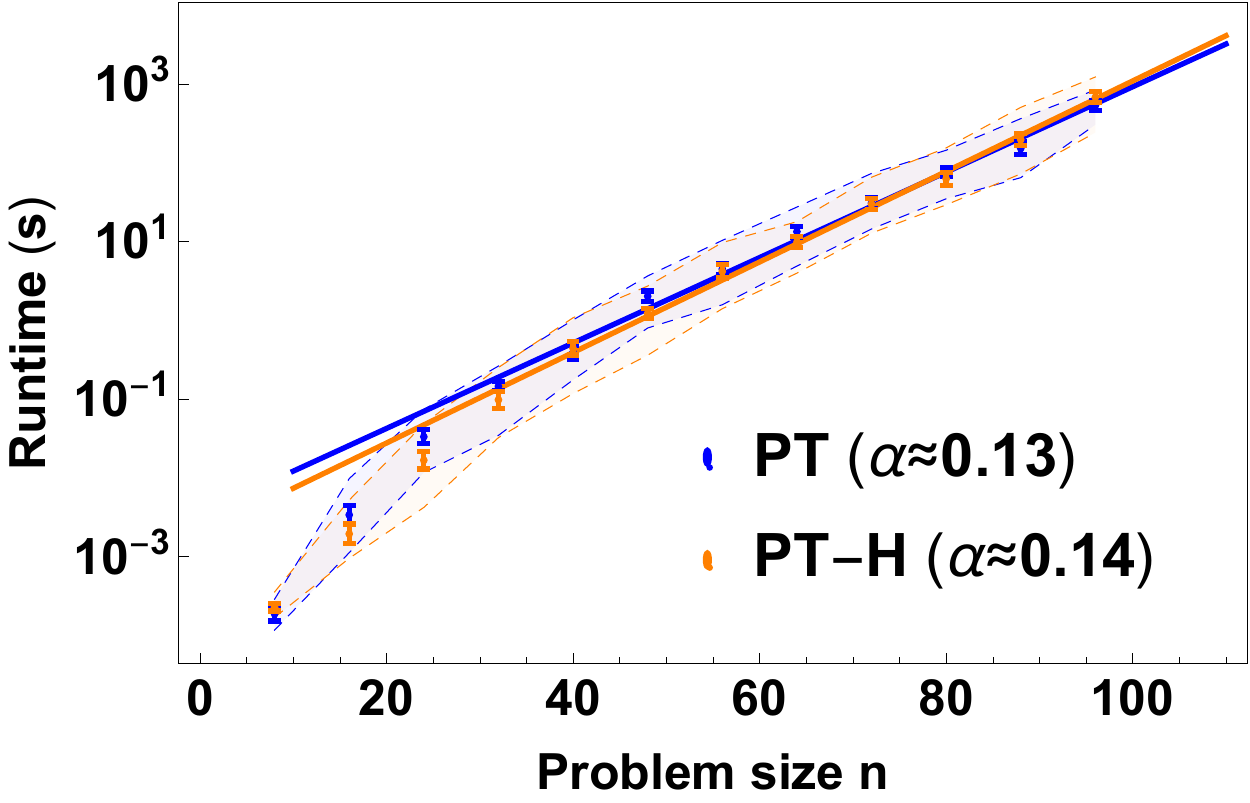}
\includegraphics[width=0.9\figurewidth]{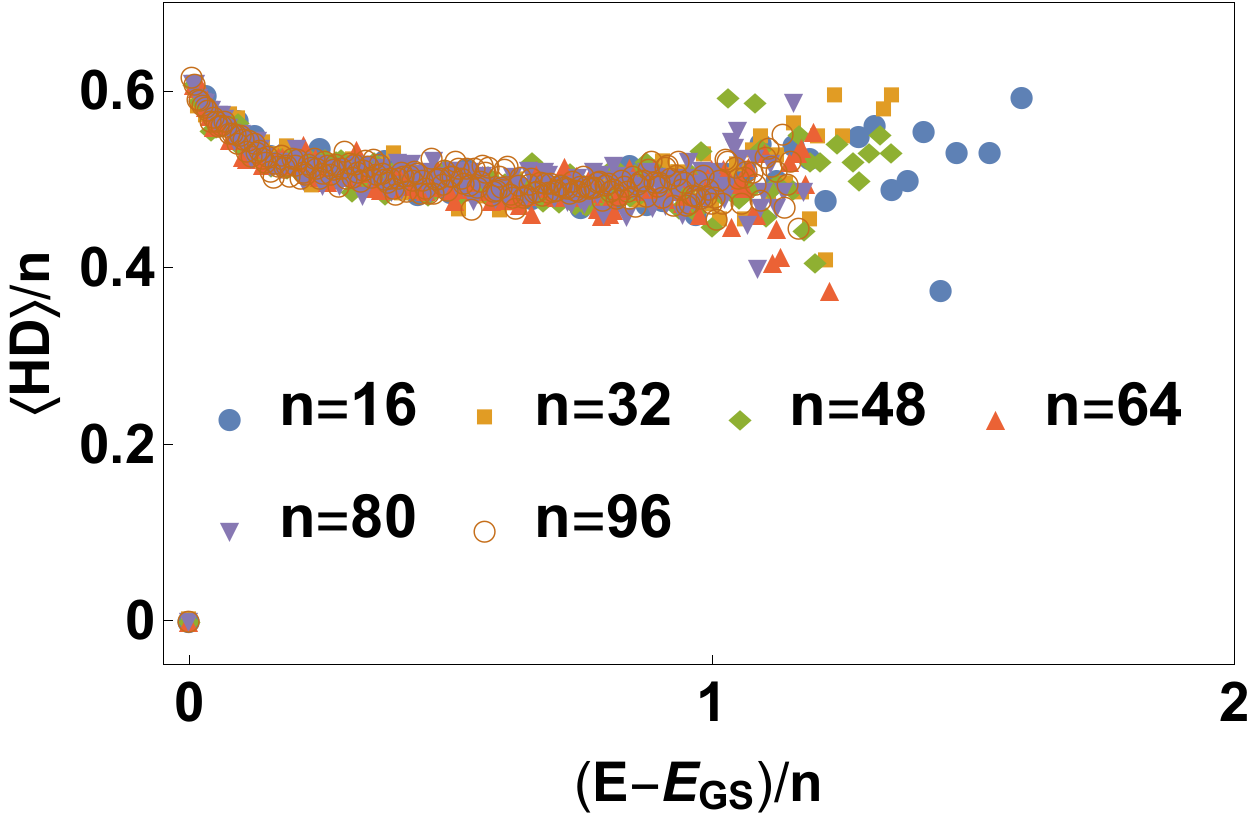}
\caption{Top: Runtime scaling of two parallel tempering variants on random 3R3X-planted Ising instances (log-linear scale). Typical runtimes (median over 100 instances) as a function of problem size is given for parallel tempering (PT) and PT augmented with Houdayer cluster moves (PT-H). Both exhibit similarly strong exponential scaling. Fits to $\e^{\alpha n}$ are shown. Error bars correspond to $1\sigma$ statistical confidence. The shaded areas correspond to the interquartile ranges for each problem size. 
Bottom: Typical Hamming distances (normalized by problem size) of local minima from the global minimum as a function of normalized residual energy. The figure demonstrates that low-lying excited states are typically $O(n)$ spin flips away from the global minimum.}
\label{fig:runtime}
\vspace{-0.7cm}
\end{center}
\end{figure}

The results are presented in Fig.~\ref{fig:runtime}(top) showing the exponential runtime scaling of both PT and PT-H  in problem size, proportional to $\e^{\alpha n}$ with $\alpha \approx 0.13$ and $0.14$, respectively, indicating a strong exponential scaling. Moreover, the Houdayer cluster updates do not seem to improve the scaling. Also shown in the figure (as shaded areas) are the interquartile ranges (the difference between 75th and 25th percentiles) of the PT and PT-H runtimes, quantitatively indicating the homogeneity in the hardness of the instances. 

To better understand why these problems, while being trivial to solve using Gaussian elimination, present severe challenges to heuristic solvers, we next examine their energy landscapes. We do so by measuring the Hamming distances between local minimum configurations, as found by the solver in the course of the simulation, and the global minimum. These are plotted in Fig.~\ref{fig:runtime}(bottom). The low-lying excited states, characterized by small yet positive residual energies, are typically $ \sim 0.6 n$ spin flips away from the global minimum, implying that reaching a local minimum is no indication of the whereabouts of the ground state. The above property is a hallmark of NP-hard problems~\cite{papadimitriou2013combinatorial}.

\begin{figure*}[ht]
\begin{center}
\includegraphics[width=0.67\figurewidth]{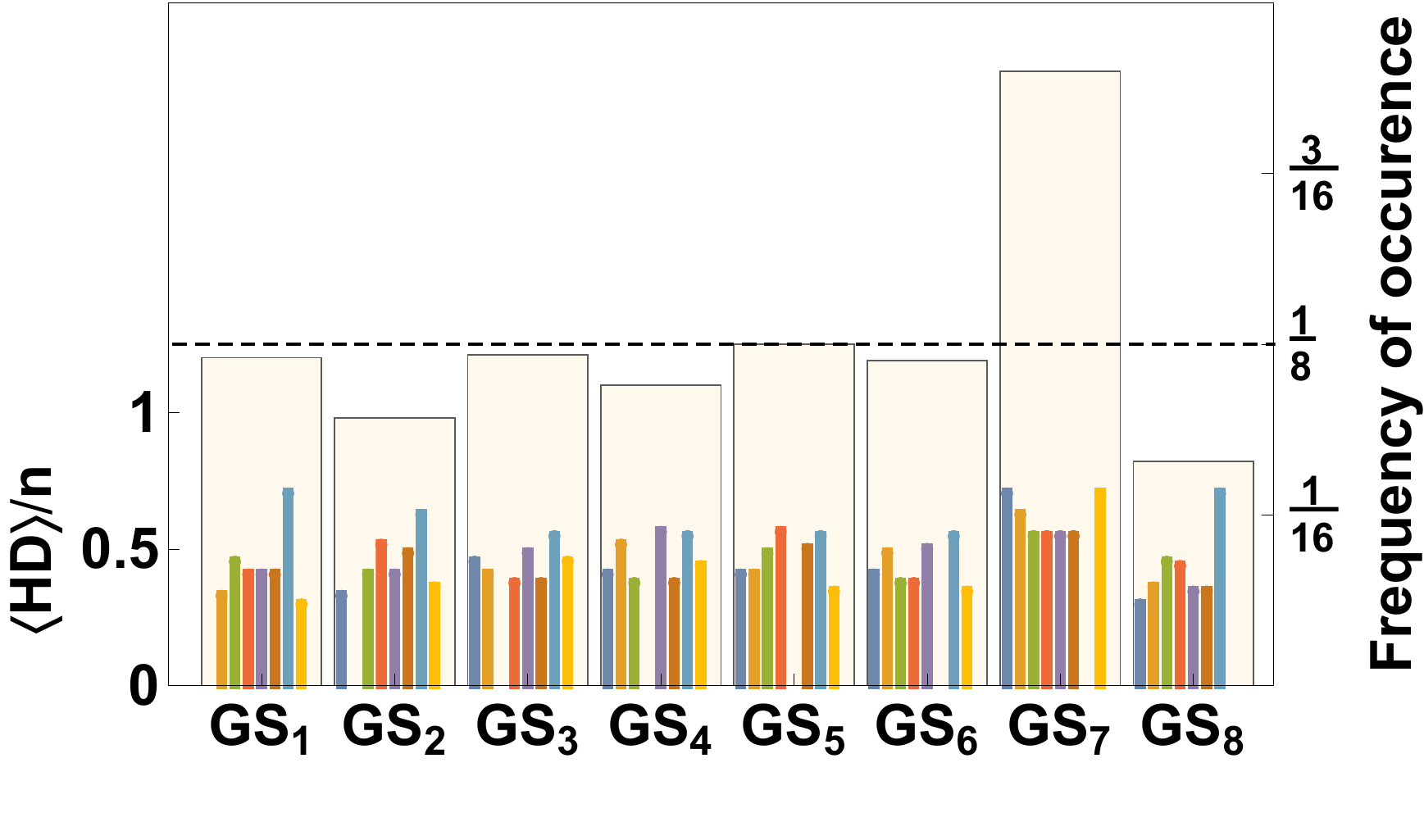}
\includegraphics[width=0.6\figurewidth]{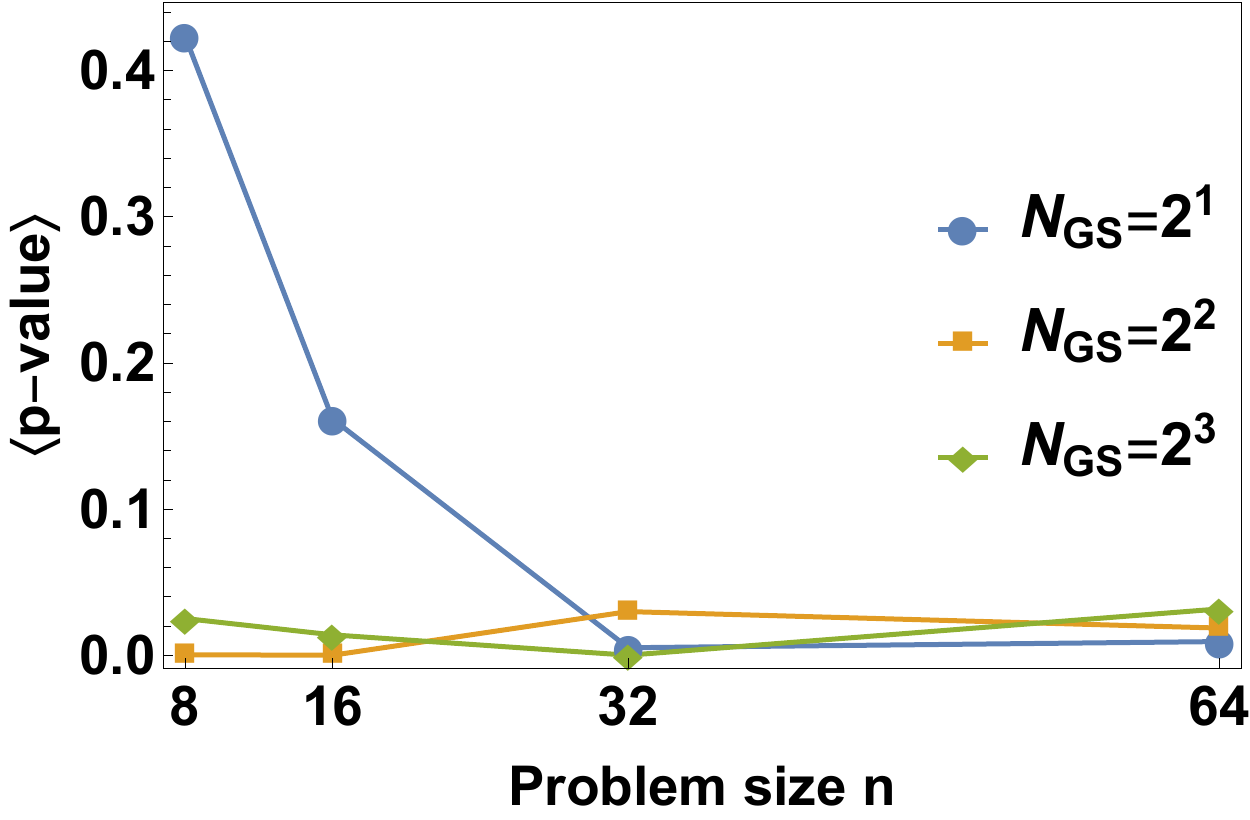}
\includegraphics[width=0.61\figurewidth]{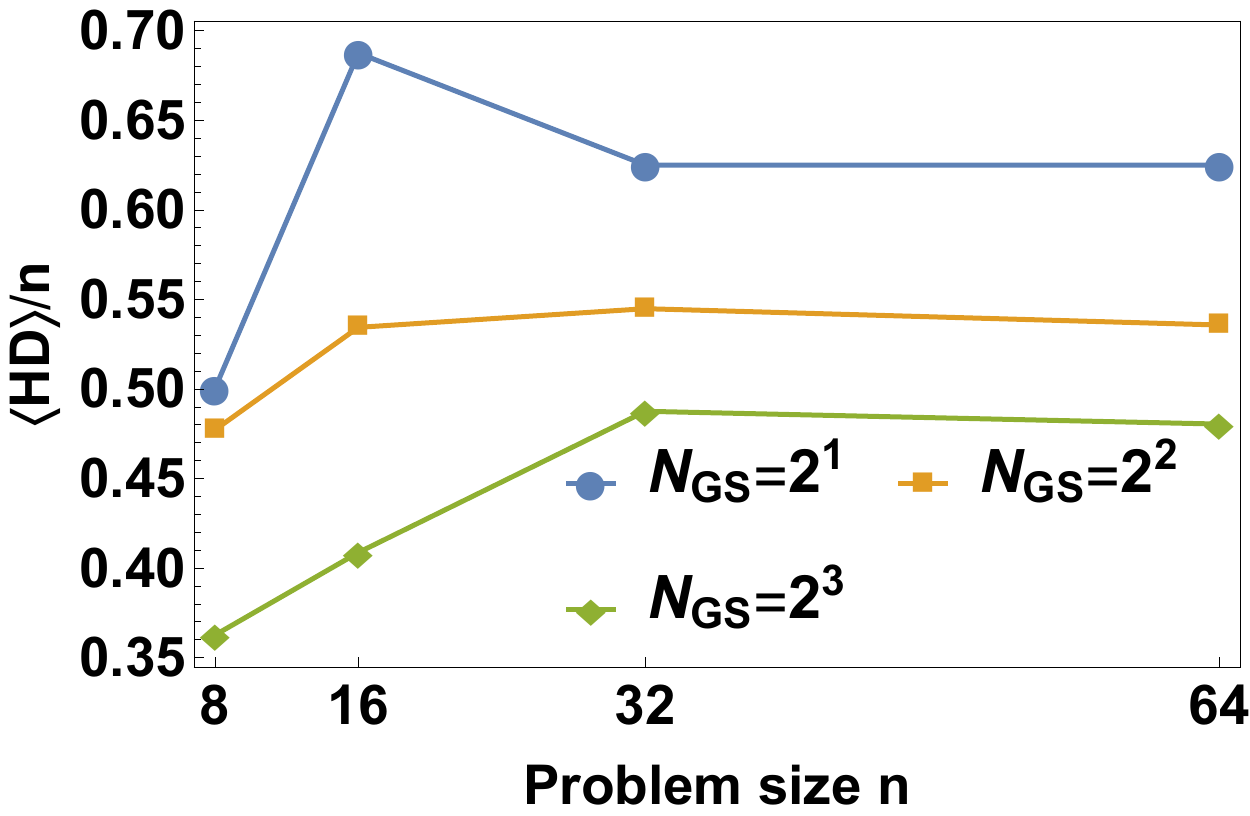}
\caption{Left: Fraction of ground-state occurrences for a random $64$-spin 3R3X instance with an 8-fold ground-state degeneracy as found by 1000 parallel tempering runs. The horizontal dashed line corresponds to uniform sampling. Also shown are the normalized Hamming distances between each ground state and all others. Middle: Typical $p$-values obtained from a one-sided $\chi^2$ test performed on the ground state occurrences of randomly generated instances with $2$-, $4$- and $8$-fold ground-state degeneracies as a function of problem size $n$. Right: Typical normalized Hamming distances between the ground states of degenerate instances  as a function of problem size $n$ for different nullities.}
\label{fig:fair}
\vspace{-0.7cm}
\end{center}
\end{figure*}


We next illustrate the usefulness of equation planting in determining the bias of Ising machines tasked with sampling the ground-state manifolds of Ising cost functions~\cite{bastea:98,martin-mayor:15,matsuda}, which in addition to optimization is considered to be one of their main suggested uses~\cite{adachi,perdomo, 2xdwave,Zhang17}.

We first construct random Ising-encoded 3R3X instances with nullities $d_N=1,2,3$ having ground-state degeneracies of $2, 4$ and $8$, respectively, at different problem sizes.  We then study the functionality of PT-H as a ground-state sampler by simply measuring the fraction of occurrences of each ground state for any given instance [see Fig.~\ref{fig:fair}(left)]. Each instance is solved 100 times, and the ground-state configurations are recorded. 

We quantify the bias of PT-H on the various instances by measuring the typical $p$-value obtained from one-sided $\chi^2$ tests performed on the tallied ground states of each instance under the hypothesis that the true distribution is uniform. The results are summarized in Fig.~\ref{fig:fair}(middle). The smallness of the $p$-values indicates PT's strong bias in sampling ground-state manifolds of these instances, implying differently sized `basins of attraction' for each ground state.
We also find that, similar to the ground state manifolds of NP-hard problems, the distances between the ground states of any given instance are typically extensive. These are depicted in Fig.~\ref{fig:fair}(right)~\footnote{The bias exhibited by PT may seem at first to contradict the Boltzmann distribution which (in accordance with the assumption of equal a priori probability)
prescribes the same probability to same-cost configurations for thermalized systems. However, we note that when PT is used as an optimizer (as it is here), the number of sweeps is not large enough for thermalization to take place.}.

\noindent{\bf Summary and conclusions.}---
We introduced a method for generating random problem sets that we argued are optimally suitable for verifiably testing the utility of Ising machines. The proposed technique allows for the construction of random problem sets that have a number of properties that are desirable for this task: (i) verified ground-state energies; (ii) controllable ground-state degeneracy; (iii) `NP-hard-like' energy landscapes; (iv) low-precision problem parameters and (v) underlying connectivity with a bounded degree.

We further illustrated the computational difficulties that heuristic state-of-the-art solvers face when tasked with finding or uniformly sampling the minimizing configurations of problem sets possessing intricate energy landscapes. We showed that equation-planted instances become difficult to solve already at small sizes, revealing rather early on the asymptotic exponential scaling expected from heuristic solvers.
In addition, we noted that, as is expected from difficult problem sets, Houdayer cluster moves~\cite{Houdayer} do not provide any additional advantages.

With the above stated, it should be noted that since the problem sets are generated from trivially solvable systems, special-purpose algorithms could conceivably be tailored to identify the underlying equations within the Ising cost function and utilize the equation structure to solve these Ising costs polynomially fast. On the other hand, the equation structure can be easily obfuscated if one (i) employs a wider range of gadgets to embed the various XORSAT clauses, (ii) constructs instances using a randomly weighted sum of terms, (iii) generates non-regular linear systems of equations with potentially different numbers of variables per equation or (iii) adds to existing XORSAT cost functions non-XORSAT gadgets whose ground states coincide with those of the instance, as these are known in advance. An interesting question that arises in this context is whether one can obfuscate such easily solvable problems in a manner that is provably intractable to `reverse engineer' back to easily solvable form. 

In the generation of the random XORSAT instances considered in this study, no specific connectivity constraints were assumed, which in turn led to the generation of Ising instances with random connections among the various spins. These instances could not be directly programmed into Ising machines that have rigid and sparse architectures~\cite{2xdwave}. This limitation may be addressed in one of two ways. The first is to convert the randomly connected instances to match the Ising machine hardware graph via embedding schemes at the price of adding auxiliary spins to the problem~\cite{embed1}. Alternatively, one can directly generate equation-planted instances that are native to the device connectivity, which may require devising novel 3-XORSAT gadgets. 

Another aspect of equation planting that has not been explored here is the relative hardness of $r$-regular $k$-XORSAT instances for different $k, r$ greater than three --- problem classes that are expected to yield yet harder instances. Varying $k, r$ would presumably allow for `hardness tunability', which is often a desired property in the benchmarking of heuristic solvers.  Another way  for achieving hardness tunability that was not pursued here is the use of non-regular random graphs, such as Poisson graphs, rather than random regular graphs~\cite{PhysRevLett.88.188701}. 
 
The NP-hard-like energy landscapes characterizing equation-planted instances coupled with the fact that their ground state manifolds can be known in advance, may also be used as a tool for gaining insight into the mechanisms underlying the operation of Ising machines. 
This is because the machines' performance on these problems serves as an indicator of the locality of the heuristic being employed. As we demonstrated above, local search approaches are not very successful whereas global approaches such as Gaussian elimination are able to decipher their inherent structure very easily. 
Another possible use of this type of problems is perhaps also the study of spin glasses, specifically the onset of temperature chaos~\cite{Tchaos1,aspelmeier:02,berthier:05}. 

One intriguing future research direction would be the development of heuristic solvers that identify XORSAT-type relations between the spins of Ising cost functions and utilize the ease with which such sub-problems can be minimized in order to efficiently find global minimal costs. 

\begin{acknowledgments}
We thank Tameem Albash and Federico Ricci-Tersenghi for useful comments and discussions. 
The research is based upon work (partially) supported by the Office of
the Director of National Intelligence (ODNI), Intelligence Advanced
Research Projects Activity (IARPA), via the U.S. Army Research Office
contract W911NF-17-C-0050. The views and conclusions contained herein are
those of the authors and should not be interpreted as necessarily
representing the official policies or endorsements, either expressed or
implied, of the ODNI, IARPA, or the U.S. Government. The U.S. Government
is authorized to reproduce and distribute reprints for Governmental
purposes, notwithstanding any copyright annotation thereon. This material is based on
research sponsored by AFRL under agreement number FA8750-18-1-0109. The
U.S. Government is authorized to reproduce and distribute reprints for Governmental
purposes notwithstanding any copyright annotation thereon.  
\end{acknowledgments}

\bibliography{refs} 

\appendix

\section{\label{app:3X3R} Generating 3-regular 3-XORSAT instances}

The simplest way to generate a uniform ensemble of $n$-spin 3-regular 3-XORSAT instances is by generating triplets of lists ${\bf v}^{(1)}, {\bf v}^{(2)}$ and ${\bf v}^{(3)}$ each of which is a random shuffling of the list $(1,\ldots, n)$.
As a next step, the 3-XORSAT clauses, or equations, are constructed from the $n$ triplets $C_i=({\bf v}^{(1)}_i,{\bf v}^{(2)}_i,{\bf v}^{(3)}_i)$ with $i=1\ldots n$. If any of the clauses are found to contain repeated indices, the process is reinitialized. The triplets of indices $C_i$ can then be used to form XORSAT equations \hbox{$x_{{\bf v}^{(1)}_i} + x_{{\bf v}^{(2)}_i} + x_{{\bf v}^{(3)}_i}= {\bf b}_i \mod 2$} where ${\bf b}$ is a randomly generated vector of $n$ bit values. These equations can also be written as \hbox{$x_{{\bf v}^{(1)}_i} \oplus x_{{\bf v}^{(2)}_i}\oplus x_{{\bf v}^{(3)}_i}= {\bf b}_i$} where $\oplus$ is the XOR operation (hence the name XORSAT). This routine ensures that every index appears in exactly three clauses, i.e., it is 3-regular. 
The degeneracy of the Ising instance generated from the above XORSAT system of equations can be found by row-reducing the system to obtain the dimension of its null space. Similarly, $r$-regular $k$-XORSAT instances can be generated for different choices of $k$ and $r$. 

\section{\label{app:gadget} Gadget construction}

The gadgets for this study were constructed using a brute-force search of the parameter space of Ising cost functions, namely sweeps overs ranges of $J_{ij}$ couplings and longitudinal fields $h_i$. While surely more efficient (and more intelligent) methods exist for constructing such gadgets, since the gadgets need only be constructed once, such methods are not particularly necessary. 
 
To construct an Ising cost function for a $k$-XORSAT equation using $a$ auxiliary spins we consider fully-connected Ising models with $k'=k+a$ spins. Beginning with $a=0$, we systematically scan through sets of integer-valued couplers and external fields (starting with $J_{i},h_i=0,\pm1,\pm 2,\ldots$).
For each choice of cost function parameters, we enumerate the ground-state configurations of the resultant model. Specifically, we test whether the ground states satisfy the XORSAT conditions, namely, that there are $2^{k-1}$ of them and that all have the same parity over the non-auxiliary spins, i.e., $\prod_{j=1}^k s_k=(-1)^b$ where $b$ is either 0 or 1.
If no choice of parameters is found, we conclude that the number of auxiliary spins $a$ is too small. We then increase $a$ by one and restart the routine.

\section{\label{app:PT} The parallel tempering simulations}
 
We outline here the technical details of the parallel tempering (PT) simulations~\cite{hukushima:96,marinari:98b} used in this study.
In PT, one considers $N_T$ independent copies of the system executing a Metropolis algorithm in parallel at different
temperatures, $T_1<T_2<\ldots< T_{N_T}$. Copies with neighboring temperatures regularly attempt to swap their
temperatures with probabilities that satisfy detailed balance~\cite{sokal:97}. In this
way, each copy performs a temperature random-walk. At high temperatures, free-energy
barriers are easily overcome, allowing for a global exploration off
configuration space. On the other hand. at lower temperatures the local
minima are explored in more detail. 

\begin{figure}[h]
\begin{center}
\includegraphics[width=0.38\textwidth]{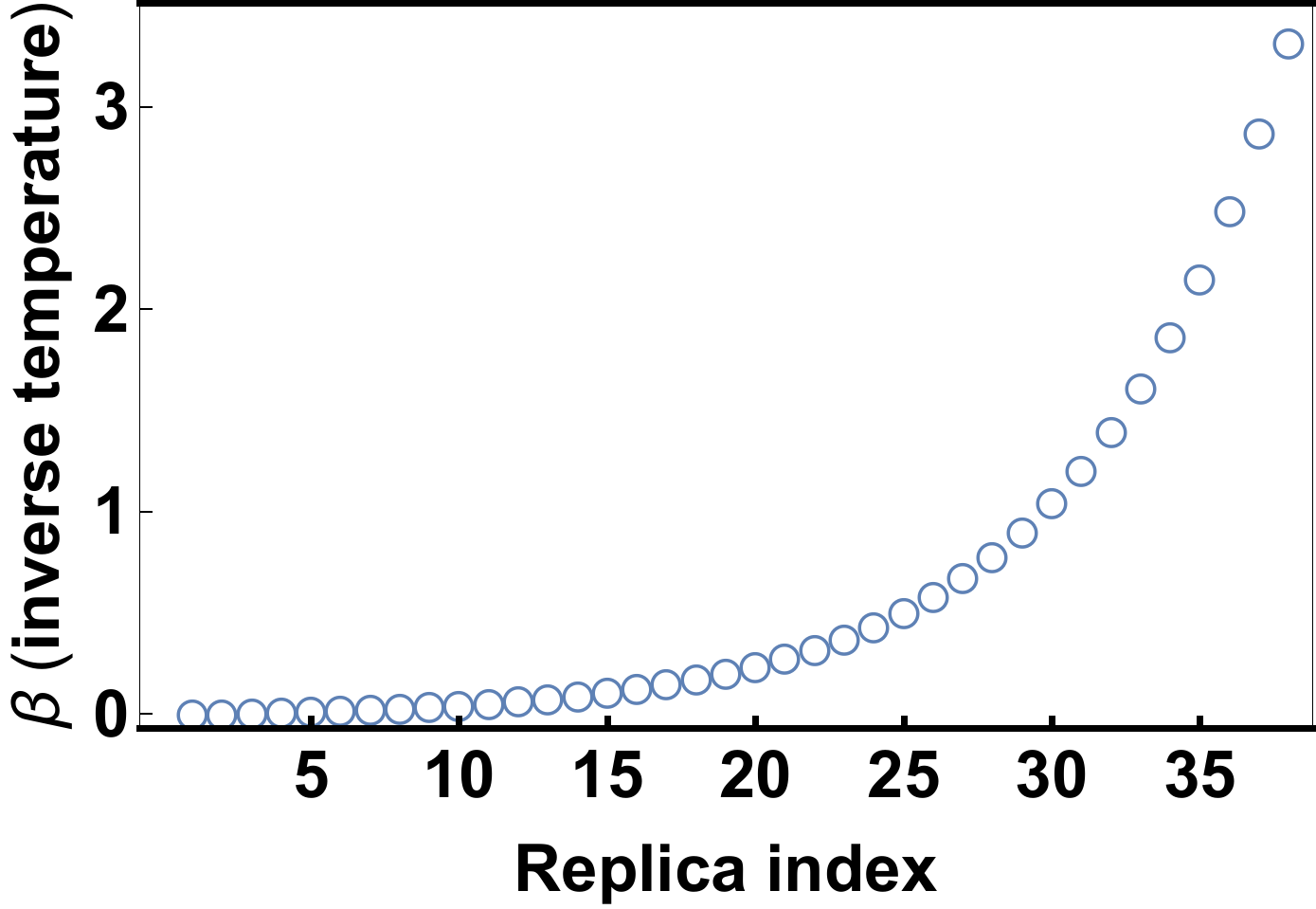}
\includegraphics[width=0.4\textwidth]{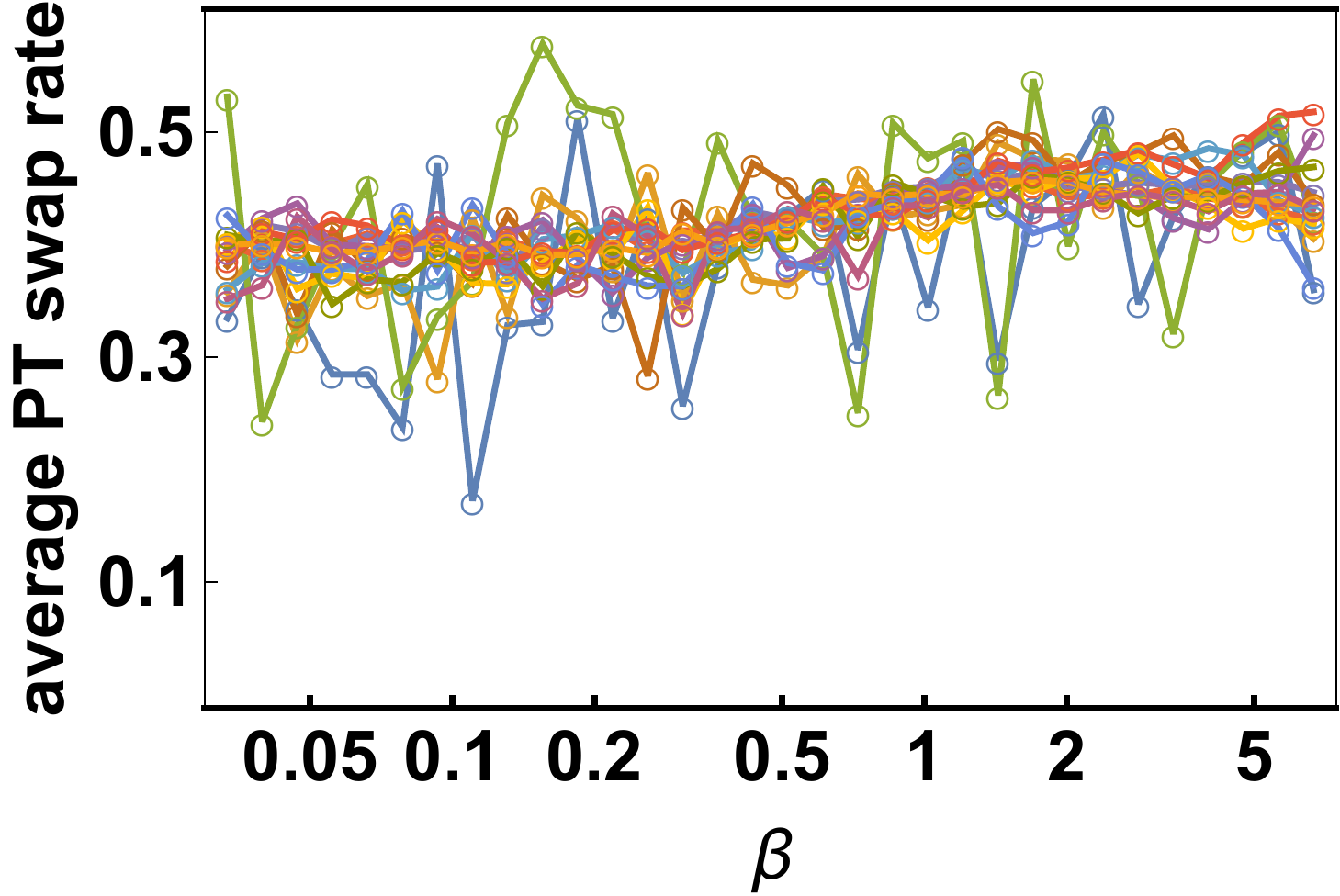}
\caption{Left: Inverse temperatures $\beta$ chosen for the parallel tempering runs. Right: Average swap rates for $10$ randomly chosen 3R3X instances. As the figure indicates, the swap rates are mainly in the range $[0.2,0.5]$.}
\label{fig:PTswapRate}
\vspace{-0.7cm}
\end{center}
\end{figure}

The temperature grid of our PT simulations consisted of $N_T=37$ inverse-temperatures $\beta$ in the range $(0.0166667, 3.33333)$ [see Fig.~\ref{fig:PTswapRate}(left)]. The maximal inverse temperature was chosen such that $\beta_{\max} |J_{\max}|=10$ to ensure an effectively zero final temperature. 
Ergodicity was maintained by the temperature-swap part of the PT.

To verify an optimized choice of the temperature mesh, the acceptance probabilities of swaps between neighboring temperatures were closely monitored to make sure that there are no bottlenecks in the flow of configurations in temperature space, i.e., that the acceptance ratios are not too small (or too high). 
Figure~\ref{fig:PTswapRate}(right) shows the swap acceptance rates for several randomly chosen instances, indicating the effectiveness of the method. 

\end{document}